# Identified charged hadron production in pp and Pb–Pb collisions with ALICE at the LHC

Maria Vasileiou[1,a] on behalf of the ALICE Collaboration

[1]University of Athens, Faculty of Physics

**Abstract.** Nuclear matter under extreme conditions can be investigated in ultra-relativistic heavy-ion collisions. The measurement of transverse momentum distributions and yields of identified particles is a fundamental step in understanding collective and thermal properties of the matter produced in such collisions. The ALICE Experiment results on identified charged hadron production are presented for pp collisions at $\sqrt{s}$ = 0.9, 2.76 and 7 TeV and for Pb–Pb collisions at $\sqrt{s_{NN}}$ = 2.76 TeV. Spectral shapes, production yields and nuclear modification factors are shown and compared to previous experiments and Monte Carlo predictions. The spectral shapes in Pb–Pb collisions indicate a strong increase of the radial flow velocity with respect to RHIC energies, which in hydrodynamic models is expected as a consequence of the increasing particle density. The observed suppression of high transverse momentum particles in central Pb–Pb collisions provides evidence for strong parton energy loss in the hot and dense medium.

## 1 Introduction

The spectra of identified particles provide a tool to study the dynamics of the quark-gluon plasma created in Pb–Pb collisions and particle production mechanisms in different momentum regions. Studying the low $p_T$ spectra ($p_T$ < 2 GeV/$c$), information about the production of soft particles from the thermalized medium can be obtained. Measurements in the high $p_T$ region ($p_T$ > 8 GeV/$c$) is important to study the hard production in the perturbative regime and jet quenching phenomena. The intermediate $p_T$ range provides information of the so-called baryon anomaly phenomena first observed at RHIC, that is the enhancement of the baryon over meson ratio as a function of $p_T$ with increasing centrality of the collision.

## 2 ALICE detector

ALICE (A Large Ion Collider Experiment) [1] is a general-purpose detector for heavy-ion studies at the CERN LHC (Large Hadron Collider). It has unique particle identification (PID) capabilities among the LHC experiments, allowing for measurements of particles in a wide range in $p_T$. The ALICE experiment consists of a combination of central-barrel detectors and several forward detector systems. The central barrel, located inside a solenoidal magnet (B = 0.5 T), covers the mid-rapidity region $|\eta| < 0.9$ over the full azimuthal angle. It includes a six-layer high-resolution Inner Tracking System (ITS), a large-volume Time Projection Chamber (TPC), electron and charged-hadron identification detectors which exploit Transition Radiation (TRD) and Time Of Flight (TOF) techniques, respectively. Small area detectors for high $p_T$ particle-identification (HMPID), photon and neutral-meson measurement (PHOS), photon and electron ID and jet reconstruction (EMCal) complement the central barrel. The forward-rapidity system includes a single-arm muon spectrometer covering the pseudorapidity range $-4.0 \leq \eta \leq -2.5$ and several smaller detectors for triggering, multiplicity measurements and centrality determination.

---

[a] Corresponding author: mvasili@phys.uoa.gr

## 3 Results

The results presented here are obtained from a data sample collected during the LHC Pb–Pb run at $\sqrt{s_{NN}} = 2.76$ TeV in 2010 and 2011, the p–Pb run at $\sqrt{s_{NN}} = 5.02$ TeV in the beginning of 2013 and the pp run at $\sqrt{s} = 2.76$ TeV in 2011.

### 3.1 Identified particle spectra and ratios

The spectra for summed charge states in central (0–5%) collisions are compared to hydrodynamical models and previous results in Au-Au collisions at $\sqrt{s_{NN}} = 200$ GeV in Fig.1 [2]. A dramatic change in spectral shapes from RHIC to LHC energies is observed, with the protons in particular showing a flatter distribution.

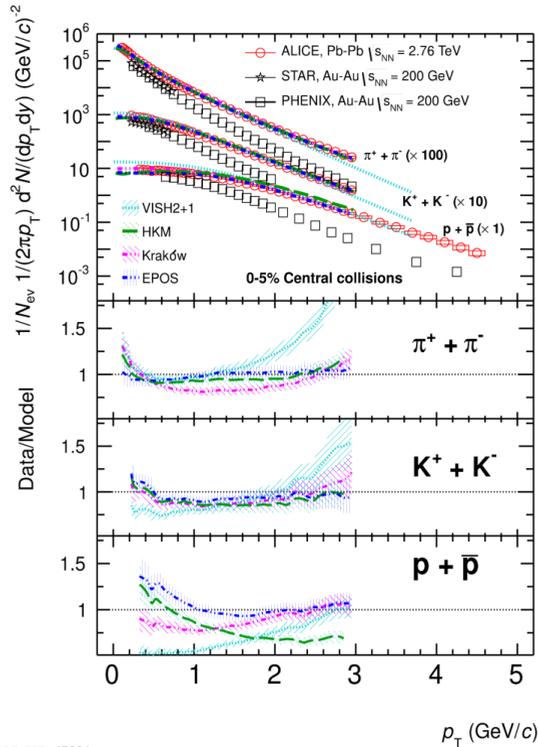

**Figure 1**. (Color online) Spectra of particles for summed charge states in the centrality bin 0–5% compared to hydrodynamical models and results from RHIC at $\sqrt{s_{NN}} = 200$ GeV. Systematic uncertainties plotted (boxes); statistical uncertainties are smaller than the symbol size.

The comparison to models based on hydrodynamics shows that VISH 2+1 [3] describes the pion and kaon spectra for $p_T < 1.5$ GeV/$c$ qualitatively well, but fails in the description of protons, which can be attributed to a lack of an explicit description of the hadronic phase. HKM [4] inserts a hadronic cascade after the hydrodynamic phase, which further transports them until final decoupling. This improves the description of the protons for $p_T < 1.0$ GeV/$c$. The Kraków [5] model provides a better description of the spectra than the other two models. This model uses temperature ($T_{ch}$) fluctuations, consequence of non-equilibrium due to the bulk viscosity. The EPOS [6] model uses breakup of flux tubes which either contribute to the bulk or escape the medium as jets. This model describes qualitatively well the data in a broad $p_T$ range. A comparison between RHIC and LHC energies based on the values of $<\beta_T>$ and $T_{kin}$ from the combined blast wave fits [7] is shown in Fig.2[2]. For central collisions, about 10% stronger radial flow is observed at the LHC with respect to RHIC. The value of $<\beta_T>$ extracted from the fit increases with centrality, while $T_{kin}$ decreases, similar to what was observed at lower energies . This is interpreted as a possible indication of a more rapid expansion with increasing centrality [7].

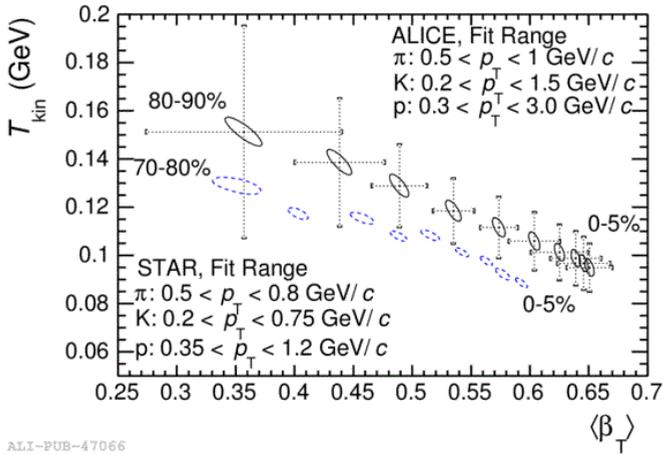

**Figure 2.** (Color online) Results of blast-wave fits compared to similar fits at RHIC energies. The uncertainty contours include the effect of the bin-by-bin systematic uncertainties and the dashed error bars represent the full systematic uncertainty. The STAR contours include only statistical uncertainties.

Fig.3 [8] shows the invariant yields measured in Pb–Pb collisions compared to those in pp collisions scaled by the number of binary collisions, $N_{coll}$ [9], obtained for the measured pp cross section [10]. For peripheral Pb–Pb collisions the shapes of the invariant yields are similar to those observed in pp collisions. For central Pb–Pb collisions, the spectra exhibit a reduction in the production of high-$p_T$ particles with respect to the reference which is characteristic of jet quenching.Fig.4 [11] shows the ratio of proton to pion spectra and kaon to pion spectra in Pb–Pb and pp collisions. For $1.5 < p_T < 5$ GeV/$c$ the Pb–Pb ratios are strongly enhanced with respect to the pp measurement. This enhancement seems to be consistent with a common velocity boost [11] (radial flow) which leads to a mass-dependent modification of the p$_T$ spectra. The kaon to pion ratio exhibits a bump at $p_T \sim 3$ GeV/c, which is qualitatively described in models including coalescence as a mechanism for parton hadronization in the low $p_T$ region. The Kraków and EPOS [12,13] models, based on a hydrodynamically expanding medium, are in better agreement with the data than the Fries [14] model, which assumes recombination as the dominant hadronization mechanism, although more recent recombination calculations also show good agreement with the measured proton to pion ratio [15]. In central collisions, the φ - meson spectrum (not shown, see [16]) is similar in shape to the proton spectrum, supporting the dominance of radial flow as the φ is a meson with a mass close to the proton

mass. At high momenta (> 10 GeV/$c$) the particle ratios in pp and Pb–Pb collisions are equal, indicating vacuum-like hadronization through fragmentation.

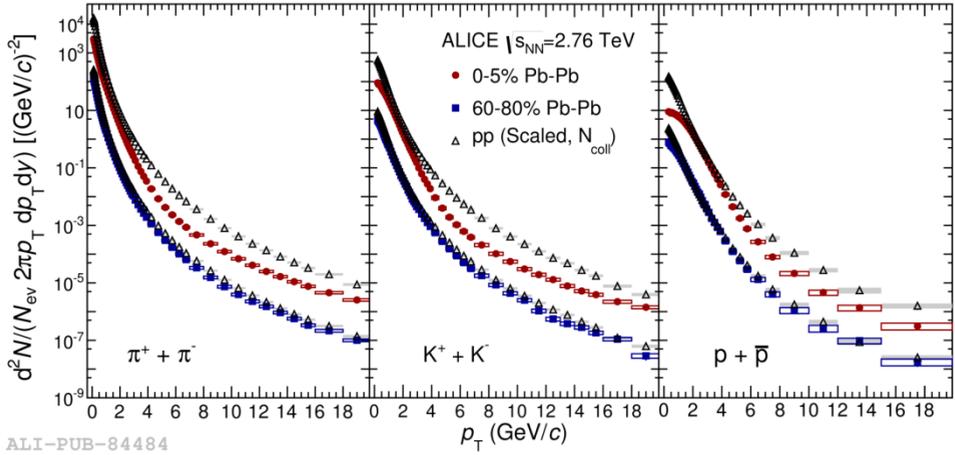

**Figure 3.** (Color online) Solid markers represent the invariant yields of identified particles in central and peripheral Pb–Pb collisions. Open points show the pp reference yields scaled by the average number of nucleon-nucleon collisions for 0-5% (upper) and 60-80% (lower). The statistical and systematic uncertainties are shown as vertical error bars and boxes, respectively.

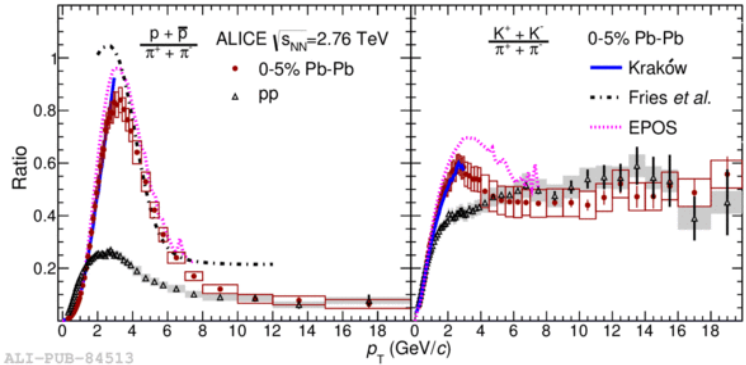

**Figure 4.** (Color online) Particle ratios as a function of $p_T$ measured in pp and the most central, 0–5%, Pb–Pb collisions. Statistical and systematic uncertainties are displayed as vertical error bars and boxes, respectively. The theoretical predictions refer to Pb–Pb collisions.

## 3.2 Nuclear modification factor

Fig.5 [11] shows the nuclear modification factor $R_{AA}$ as a function of $p_T$, defined as the ratio of the Pb–Pb spectra to the $N_{coll}$ scaled pp spectra shown in Fig.3. The $R_{AA}$ for the sum of kaons and protons is included as it allows a quantitative comparison to the $R_{AA}$ of pions. For $p_T < 10$ GeV/$c$ protons appear to be less suppressed than kaons and pions, consistent with the particle ratios shown in Fig.4. At larger $p_T$ (> 10 GeV/$c$) all particle species are equally suppressed; so despite the strong energy loss observed in the most central heavy-ion collisions, the particle composition and ratios at high $p_T$ are similar to those in vacuum. In Fig. 6 [17] the measurement of the nuclear modification factor in p–Pb collisions

$R_{\text{pPb}}$ is compared to that in central (0-5% centrality) and peripheral (70-80%) Pb–Pb collisions $R_{\text{PbPb}}$. Proton-lead collisions provide a control experiment to clearly establish whether the initial state of the colliding nuclei plays a role in the observed high-$p_T$ hadron production in Pb–Pb collisions. $R_{\text{pPb}}$ is observed to be consistent with unity for transverse momenta higher than about 2 GeV/$c$. This important measurement demonstrates that the strong suppression observed in central Pb–Pb collisions at the LHC is not due to an initial-state effect, but it is a final state effect related to the hot matter created in high-energy heavy-ion collisions.

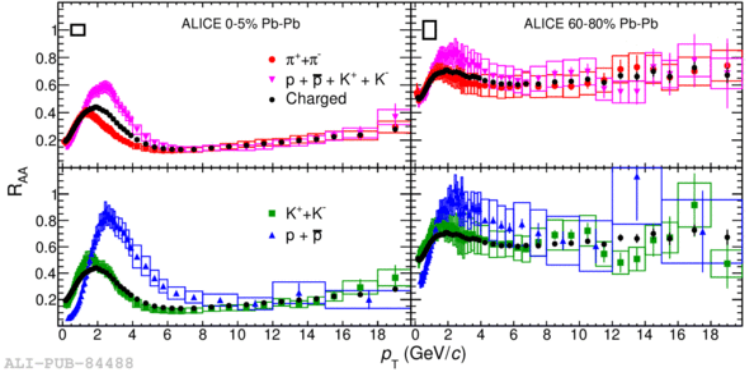

**Figure 5.** (Color online) The nuclear modification factor $R_{AA}$ as a function of $p_T$ for different particle species. Results for 0–5% (left) and 60–80% (right) collision centralities are shown. Statistical and systematic uncertainties are plotted as vertical error bars and boxes around the points, respectively. The total normalization uncertainty (pp and Pb–Pb) is indicated by the black boxes in the top panels [19].

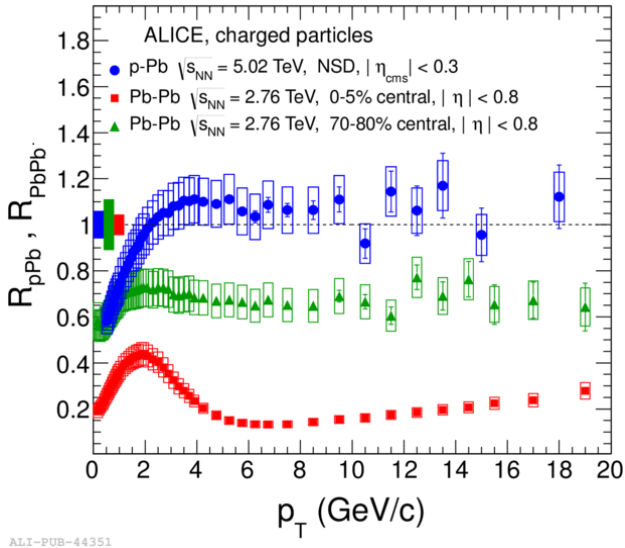

**Figure 6.** (Color online) The nuclear modification factor of charged particles as a function of transverse momentum in p–Pb collisions at $\sqrt{s_{NN}}$ = 5.02 TeV compared to measurements in central (0-5%) and peripheral (70-80%) Pb–Pb collisions at $\sqrt{s_{NN}}$ = 2.76 TeV.

In Fig. 7 [18] we compare the measurement of the nuclear modification factor for inclusive primary charged-particle ($h^{\pm}$) production in p–Pb collisions to that in central (0–5 % centrality) Pb–Pb collisions measured by ALICE [19] and CMS [20]. The ALICE p–Pb data show no sign of nuclear matter modification of hadron production at high $p_T$ and are fully consistent with the observation of binary collision scaling in Pb–Pb of observables which are not affected by hot QCD matter (direct photons [21] and vector bosons [22, 23]).

## 4 Conclusions

The spectra of identified charged particles measured by ALICE in pp and Pb–Pb collisions have been presented. Central Pb–Pb collisions spectral shapes indicate a strong increase of the radial flow velocity with $\sqrt{s_{NN}}$, which in hydrodynamic models is expected as a consequence of the increasing particle density. Strong high-$p_T$ production suppression is measured in central Pb–Pb collisions. No suppression is observed at high-$p_T$ in minimum bias p–Pb collisions. So, the strong suppression of high-$p_T$ hadrons observed in central Pb–Pb collisions is not due to an initial state effect but may instead be a signature of the hot matter created in Pb–Pb collisions.

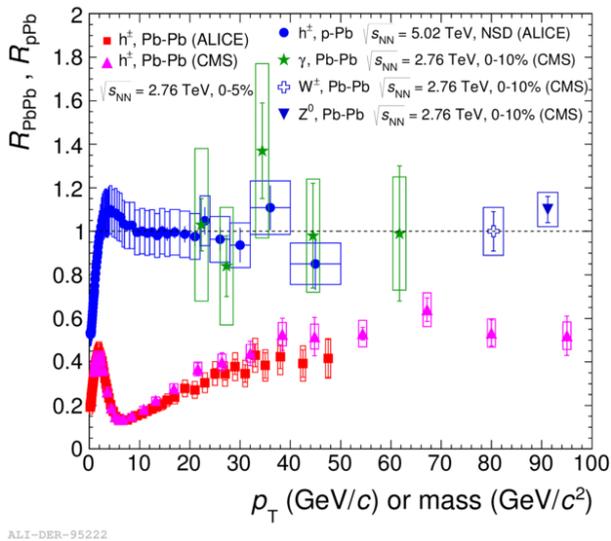

**Figure 7**. (Color online) Transverse momentum dependence of the nuclear modification factor $R_{pPb}$ of charged particles ($h\pm$) measured in minimum-bias (NSD) p–Pb collisions at $\sqrt{s_{NN}}$ = 5.02 TeV in comparison to data on the nuclear modification factor $R_{pPb}$ in central Pb–Pb collisions at $\sqrt{s_{NN}}$ = 2.76 TeV. The Pb–Pb data are for charged particle [19, 20], direct photon [21], $Z_0$ [22] and $W_{\pm}$ [23] production. All data are for midrapidity.